\documentclass{aa}
\usepackage{txfonts}
\input epsf
\usepackage{graphicx}
\usepackage{longtable}
\usepackage[mathcal]{eucal}
\usepackage{amssymb}
\usepackage{amsmath}
\usepackage{natbib}
\usepackage[T1]{fontenc}
\usepackage{color}

\def\kms {km\,s$^{-1}$}

\begin{document}

\title{Evolution of the magnetic field of \object{Betelgeuse} from 2009 -- 2017
\thanks{Based on observations  obtained at the T\'elescope Bernard Lyot (TBL) at Observatoire du Pic du Midi, CNRS/INSU and Universit\'e de Toulouse, France, and at the Canada-France-Hawaii Telescope (CFHT) which is operated by the National Research Council of Canada, CNRS/INSU and the University of Hawaii.}}

\author{P. Mathias\inst{1}, M. Auri\`ere\inst{1}, A. L\'opez Ariste\inst{2}, P. Petit\inst{2}, B. Tessore\inst{3}, E. Josselin\inst{1,3}, A. L\`ebre\inst{3},
J. Morin\inst{3}, G. Wade\inst{4}, \\
F. Herpin\inst{5}, A. Chiavassa\inst{6}, M. Montarg\`es \inst{7}, R. Konstantinova-Antova\inst{8}, \\
P. Kervella\inst{9}, G. Perrin\inst{9}, J.-F. Donati\inst{2}, J. Grunhut\inst{10}}


\institute{
IRAP, Universit\'e de Toulouse, CNRS, UPS, CNES, 57 avenue d'Azereix, 65000, Tarbes, France
\and
IRAP, Universit\'e de Toulouse, CNRS, UPS, CNES, 14 Avenue Edouard Belin, 31400 Toulouse, France
\and
Universit\'e de Montpellier, CNRS, LUPM, Place Eug\`ene Bataillon, 34095 Montpellier, France
\and
Dunlap Institute for Astronomy and Astrophysics, University of Toronto, Rm 101, 50 St. George Street, Toronto, ON M5S 3H4, Canada
\and
Laboratoire d'Astrophysique de Bordeaux, Univ. Bordeaux, CNRS, B18N, Allée Geoffroy Saint-Hilaire, 33615, Pessac, France
\and
Universit\'e C\^ote d'Azur, Observatoire de la C\^ote d'Azur, CNRS, Lagrange, CS 34229, 06304, Nice Cedex 4, France
\and
Institute of Astronomy, KU Leuven, Celestijnenlaan 200D B2401, 3001 Leuven, Belgium
\and
Institute of Astronomy and NAO, Bulgarian Academy of Sciences, 72 Tsarigradsko shose, 1784, Sofia, Bulgaria
\and
LESIA, Observatoire de Paris, PSL Research University, CNRS, UPMC, Univ. Paris-Diderot, 5 place Jules Janssen, 92195, Meudon, France
\and
Department of Physics, Royal Military College of Canada, PO Box 17000, Station Forces, Kingston, ON, K7K 4B4, Canada}

 \date{Received 2017 December 26; accepted 2018 March 15}

\abstract
{\object{Betelgeuse} is an M-type supergiant that presents a circularly polarized (Stokes $V$) signal in its line profiles, interpreted 
in terms of a surface magnetic field.}
{The weak circular polarization signal has been monitored over 7.5\,years in order to follow its evolution on different
timescales, and eventually to determine its physical origin. 
Linear polarization measurements have also been obtained regularly in the last few years.}
{We used both the ESPaDOnS and Narval spectropolarimeters to obtain high signal-to-noise ratio (S/N) spectra, which were processed by means of the
least-squares deconvolution (LSD) method. 
In order to ensure the reality of the very weak circular polarization, special care has been taken to limit instrumental effects.
In addition, several tests were performed on the Stokes $V$ signal to establish its stellar and Zeeman origin.}
{We confirm the magnetic nature of the circular polarization, pointing to a surface magnetic field of the order of 1\,G.
The Stokes $V$ profiles present variations over different timescales, the most prominent one
being close to the long secondary period (LSP; around 2000\,d for Betelgeuse) often invoked in red evolved stars.
This long period is also dominant for all the other Stokes parameters.
The circular polarization is tentatively modeled by means of magnetic field concentrations mimicking spots, showing in particular that 
the velocity associated with each ``spot'' also follows the long timescale, and that this signal is nearly always slightly redshifted.}
{From the coupled variations of both linear and circular polarization signatures in amplitude, velocity and timescale, we favour 
giant convection cells as the main engine at the origin of polarization signatures and variations in all the Stokes parameters. 
This strengthens support for the hypothesis that large convective cells are at the origin of the LSP.}

   \authorrunning {P. Mathias et al.}
   \titlerunning {The magnetic field of Betelgeuse}

\maketitle

\section{Introduction}

Cool evolved stars play a major role in the enrichment of the interstellar medium through their strong winds.
However, the mechanisms that drive mass loss from these stars are not well understood. 
Mechanisms that are often invoked include thermal gas and radiation pressure, acoustic and shock waves, Alfv\'en waves, magnetism, 
and most probably other additional phenomena.
Magnetism is one of these factors, and in this context, dedicated spectropolarimetric studies of cool evolved 
stars have been undertaken. Recent examples include the Mira variable star $\chi$\,Cyg \citep{laf14}, the radially pulsating RV\,Tauri stars 
\citep{swl15}, some red giants \citep{akc15} as well as FGK supergiants \citep{gwh10} which suggest that magnetic 
fields may well be present in all cool, evolved stars.

One of the first detections of the magnetic field of a red supergiant (RSG) concerned \object{Betelgeuse}: using the Narval spectropolarimeter,
\citet{adk10} detected a weak circular polarization Stokes $V$ signal.
From the complex behavior of the Stokes $V$ profile, and using the center of gravity technique \citep{rs79}, the 
longitudinal magnetic field integrated over the stellar disk was estimated to be of the order of 1\,G \citep{adk10}.
In addition, it was also noted \citep{pak13} that the double-peaked Stokes $V$ profiles of \object{Betelgeuse} all possess a significant 
level of asymmetry, and are additionally red-shifted by about 9\,\kms\ with respect to the Stokes $I$ profiles.
Finally, a recent study \citep{tlm17} also reports the magnetism of two other RSGs, CE\,Tau and $\mu$\,Cep, which also present 
complex structures within their Stokes $V$ signal.

The distorted Stokes $V$ signatures often observed in RSGs suggest topologically complex magnetic fields, presumably generated by dynamo
action.
The engine would be related to either giant cells \citep{sl71} or to the supergranulation cells of \citet{s75}.
These gigantic convection cells have been reproduced through numerical simulations \citep[e.g.,][]{fsd02,cfm11a} and directly seen 
in spatially resolved observations of for example, Betelgeuse \citep{hpl09}.
These flows of plasma could then create global and local magnetic fields in RSGs \citep{df03}.

Generally speaking, atmospheric motions in RSGs are difficult to understand \citep[e.g.,][]{jp07}, and this complex velocity field is 
probably the main cause of line profile broadening, of the order of 20\,\kms.
In addition, RSGs often show complex photometric and spectroscopic variability on several timescales, ranging from hundreds to thousands days \citep[e.g.,][]{pk14}.
In particular, \object{Betelgeuse} has at least two photometric timescales, a ``short'' one, of the order of 400\,d, and a longer one,
often called the long secondary period (LSP), of the order of 2000\,d \citep{ksb06}.

In order to extend the pioneering study of \citet{adk10}, dedicated to the first detection and the study of the magnetic field at the surface of Betelgeuse, 
we undertook a (still ongoing) spectropolarimetric follow-up of \object{Betelgeuse}.
Stokes $V$ observations are described in Sect.\,2, and their stellar origin is established in Sect.\,3.
The $V$ signal is then analyzed in terms of magnetic field in Sect.\,4.
A rough frequency analysis of the variations of the Stokes parameters is presented in Sect.\,5. 
Then, an attempt to model the Stokes parameters in terms of bright spots and magnetic concentrations is described in Sect.\,6.
Finally, some concluding remarks are discussed in Sect.\,7.

\section{Observations with Narval and ESPaDOnS}

We carried out long-term spectropolarimetric monitoring of \object{Betelgeuse} from September 2009 to April 2017 using Narval \citep{a03} and 
ESPaDOnS \citep{dcl06}, representing eight seasons of data acquired on 76 dates. 
We note that from November 2013, linear polarization Stokes $Q$ and $U$ observations have also been obtained, with Narval only, 
quasi-simultaneously with each Stokes $V$ series.
All spectra have a resolving power of 65\,000 and cover the wavelength range $370-1048$\,nm.

A standard polarization observation consists of a series of 4 sub-exposures, following the procedure described by \citet{sdr93}. 
To avoid saturation of the CCD detector, we performed very short exposures (3-5\,s for Narval, depending on sky quality, and 1\,s for ESPaDOnS, for each sub-exposure). 
We obtained around ten Stokes $V/I_c$ series per observing night, which are averaged. 
For the Stokes parameters $Q/I_c$ and $U/I_c$ obtained with Narval, because the signal amplitude is larger by a factor of about ten compared to the Stokes $V$, 
the number of exposures was reduced.
Also included in the measurements are the ``diagnostic null'' spectra $N_{1,2}$ obtained from different combinations of the four sub-exposures. 
These spectra are in principle featureless, and are used to diagnose the potential presence of spurious contributions to the Stokes $V$ spectrum. 
Each single spectrum used in this work has a peak signal-to-noise ratio (S/N) ranging from 1700 to 2100 in Stokes $I$ (per 1.8\,\kms\ spectral bin). 
Details of the observing and reduction procedure are described by \citet{dsc97}, \citet{awk09} and \citet{alm16}.
We note here that all spectra presented in the different figures in this paper have been smoothed for clarity using a moving average over three pixels;
the analysis was however performed using unsmoothed data.
Table\,\ref{tab1} presents the journal of observations giving the date and heliocentric Julian day corresponding to mid-observation, the
instrument (N for Narval, E for ESPaDOnS), the number of Stokes $V$ spectra collected on the given night, a unique label used to indicate each night
(or the average over several close nights), and the season number (starting from autumn 2009), defined as the epoch when Betelgeuse is 
observable (typically September -- April).
Table\,\ref{tab2} is the journal of the Stokes $Q$ and $U$ observations, extending the initial monitoring already 
presented in \citet{alm16}, giving the date of the observation night as well as the characteristics of the spots \citep[Sect.\,6, see also][]{alm16}.
To obtain a high-precision diagnosis of the spectral line polarization, the least-squares deconvolution method \citep[LSD,][]{dsc97}
was applied to each reduced Stokes $I$, $Q$, $U,$ and $V$ spectrum.
We used a solar abundance line mask \citep[similar to that of][]{alm16}, calculated from data provided by VALD \citep{kpr99} for an effective 
temperature of 3750\,K, $\log g =0.0$, and a microturbulence of 4.0\,\kms, consistent with the physical parameters of \object{Betelgeuse} 
\citep{jp07,lbh84}. 
The mask contains about 15\,000 atomic lines with a depth larger than 40\,\% of the continuum.
Application of LSD using this mask allows the detection of clear polarisation structures in the LSD profiles that are discussed below.

\setcounter{table}{1}
\begin{table*}
\caption{Log of Narval (Stokes $U$ and $Q$) observations of Betelgeuse and polarimetric measurements (see Sect.\,6).
We note that previous $QU$ measurements are summarized by \citet{alm16}.}
\label{tab2}
\centering
\begin{tabular}{l c l c c c c c c c c}
\hline\hline
Date             & Stokes      & ${P_L}_1$  & ${\theta}_1$ & ${\chi}_1$ & ${\mu}_1$  & ${P_L}_2$  & ${\theta}_2$    & ${\chi}_2$ & ${\mu}_2$ \\
                 &             & 10$^{-4}$  &  $^\circ$    &  $^\circ$  & $^\circ$   &  10$^{-4}$ & $^\circ$        & $^\circ$   & $^\circ$  \\
\hline
18-19 September 2015 & $8Q 8U$ &  2.6       & 48.0         &  138.0     & 77.3       & 2.6        &  120.5          & 210.5      &  77.7     \\
16 October 2015      & $8Q 8U$ &  2.7       & 34.6         &  124.9     & 73.0       & 3.0        &  121.2          & 211.2      &  77.7     \\
09 December 2015     & $8Q 8U$ &  2.4       & 78.1         &  168.1     & 83.6       & 1.7        &  133.1          & 223.1      &  77.8     \\
20-21 January 2016   & $8Q 8U$ &  1.6       & 90.3         &  180.3     & 81.5       & 1.6        &  145.6          & 235.6      &  77.8     \\
16 February 2016     & $8Q 8U$ &  0.5       & 87.5         &  177.5     & 79.4       & 0.2        &  162.3          & 252.3      &  82.0     \\
12 March 2016        & $4Q 4U$ &  2.6       & 28.7         &  118.7     & 77.3       & 0.8        &   93.2          & 183.2      &  64.7     \\
06 April 2016        & $4Q 4U$ &  2.4       & 17.9         &  107.9     & 70.9       & 0.8        &  107.1          & 197.1      &  66.9     \\
11 September 2016    & $4Q 4U$ &  1.8       & 25.2         &  115.2     & 73.0       & 1.7        &  106.3          & 196.3      &  60.0     \\
08 October 2016      & $4Q 4U$ &  2.7       & 17.7         &  107.7     & 73.0       & 1.8        &  108.9          & 198.9      &  62.4     \\
01 November 2016     & $4Q 4U$ &  2.6       & 18.2         &  108.2     & 70.9       & 1.8        &   99.0          & 189.0      &  71.3     \\
03 December 2016     & $4Q 4U$ &  2.5       & 45.0         &  135.0     & 75.2       & 1.6        &  102.3          & 192.3      &  73.5     \\
18 December 2016     & $4Q 4U$ &  2.6       & 52.0         &  142.0     & 75.2       & 1.6        &  115.7          & 205.7      &  75.6     \\
17 February 2017     & $4Q 4U$ &  1.7       &106.2         &  196.2     & 79.4       & 1.1        &    7.1          &  97.1      &  73.5     \\
03 April 2017        & $2Q 2U$ &  1.4       & 24.7         &  114.7     & 70.9       & 1.2        &  178.4          & 268.4      &  75.6     \\
11 April 2017        & $2Q 2U$ &  1.8       & 18.4         &  108.4     & 70.9       & 0.9        &  176.6          & 266.6      &  71.3     \\
17 April 2017        & $2Q 2U$ &  1.7       & 18.2         &  108.2     & 70.9       & 0.8        &  175.2          & 265.2      &  71.3     \\
\hline
\hline
\end{tabular}
\tablefoot{Columns give the date, the number of Stokes $QU$ obtained, then for spot1 and spot2, observed maximum of linear polarization $P_L$, polarization
angle $\theta$, position angle $\chi$, and angle to center $\mu$.}
\end{table*}

\section{The stellar origin of the Stokes $V$ signal}

A spectropolarimetric survey of supergiants performed with ESPaDOnS \citep{gwh10} obtained a marginal detection for \object{Betelgeuse}.
This marginal detection was immediately confirmed by \citet{adk10} who detected the magnetic field of Betelgeuse during the spring of 2010 
using the Narval spectropolarimeter.
As described in Sect.\,2, in order to obtain a high-precision diagnosis from the Stokes parameters, both teams used the LSD method \citep{dsc97}. 
\citet{adk10} give arguments showing that the detected Stokes $V$ signal is not spurious. 
Since this discovery, \object{Betelgeuse} has been followed up during each visibility season with Narval or ESPaDOnS.
The Stokes $V$ parameter is found to have a variable amplitude and shape, with a strength remaining at the level of a few times 10$^{-5}$ 
of the unpolarized continuum. 
However, the null polarization signals $N_{1,2}$ sometimes present features that can potentially be diagnostics of problems with the circular polarization 
analysis.
Moreover, since linear polarization (which is about ten times larger than the circular polarization for Betelgeuse) has been detected \citep{alm16}, it may also lead 
to crosstalk into the Stokes $V$ spectra \citep[as shown by][]{tlm17}, again preventing a clear interpretation of this latter signal.

Therefore, the reality of the $V$ signal should first be carefully evaluated in order to interpret our relatively low-amplitude signal as a time series.
It should be pointed out that, while the Zeeman signature of Betelgeuse is weak, weaker magnetic detections have been obtained both in cool and tepid bright stars, for example, other supergiants 
\citep{gwh10}, red giants \citep[Pollux, Aldebaran, Arcturus and Alphard, ][]{akc15}, and the A-type stars Vega \citep{lpb09}, Sirius \citep{pla11}, 
$\beta$\,Leo, and $\theta$\,Leo \citep{bpl16}. 
For all these detections, the polarization origin of the Stokes $V$ signal has been evaluated, and a Zeeman origin is the most likely hypothesis. 
More recently, \citet{tlm17} have detected magnetic fields at the surface of two RSGs (\object{CE\,Tau} and \object{$\mu$\,Cep}), 
and clarified the nature of the spurious $N$ signal that can appear in observations of RSGs.

\subsection{Influence of crosstalk on Stokes $V$ measurements}

The twin spectropolarimeters ESPaDOnS and Narval experience reciprocal crosstalk between linear and circular polarization which must be 
taken into account when investigating very small linear polarization levels in the presence of very high circular polarization signals, and vice versa \citep[e.g.,][]{swk12}.
Since linear polarization has been detected in the spectral lines of \object{Betelgeuse} \citep{alm16}, it is important to further study the crosstalk from (strong) 
linear to (weak) circular polarization signals.

In the case of ESPaDOnS, a deep investigation of the crosstalk problem was carried out, and the crosstalk was ultimately reduced below the 1\,\% level at the time of our 
observations \citep{bbs10,swk12}, namely at 0.5\,\% or lower for both Stokes $Q$ or $U$ to Stokes $V$.

The crosstalk of Narval has been measured directly on the sky by observing the (strongly) magnetic Ap star $\gamma$\,Equ.
In September 2009 \citep{swk12}, it was found to be 3.1\,\% from Stokes $V$ to Stokes $Q$ and 
below 0.2\,\% from Stokes $V$ to Stokes $U$, the process being supposed to be reciprocal (and assuming no crosstalk between $Q$ and $U$).
In September 2016, the same test provided values of 1\,\% and 1.5\,\% respectively.
Thus, crosstalk of Narval from $Q$ to $V$ and $U$ to $V$, assuming that it is reciprocal, is at most of the order of a few percent (3\,\%).

Recent Narval observations of the RSG \object{$\mu$\,Cep} by \citet{tlm17} show that this star presents, as does \object{Betelgeuse}, 
a much more important linear polarization signal than circular polarization.
From a dedicated observational procedure performed with Narval, these authors are able to disentangle and model the crosstalk.
For $Q$ to $V$ and $U$ to $V$, they obtained respectively 3.6\,\% and 1.4\,\%, a result consistent with that obtained above for the 
Ap star, and that confirms the reciprocity between linear and circular polarization signals.

\begin{figure}[h]
\centering
\includegraphics[width=7 cm,angle=-90] {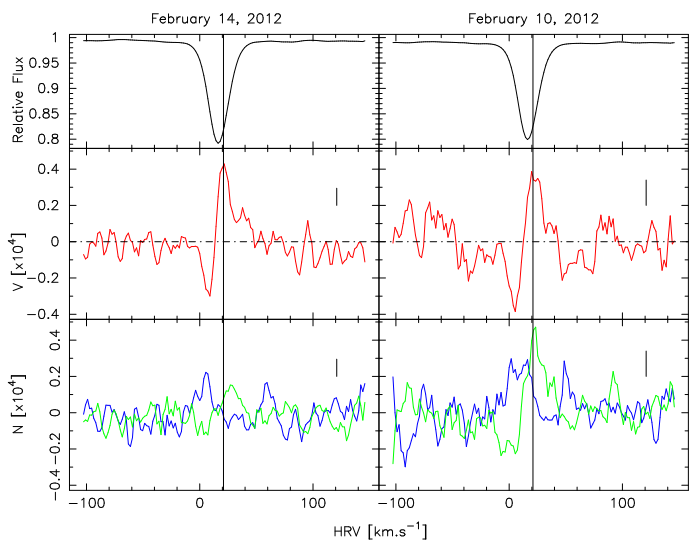} 
\caption{Stokes $I$, Stokes $V$, and null polarization $N_{1}$ (blue) and $N_{2}$ (green) LSD profiles of \object{Betelgeuse} for February 14 
(ESPaDOnS) and 10 (Narval), 2012. 
Small vertical lines represent typical error bars.
The vertical line corresponds to the heliocentric radial velocity (HRV) of Betelgeuse (about 21\,\kms).}
\label{n1n2}
\end{figure}

For Betelgeuse, Stokes $Q$ and $U$ observations with Narval present signals up to the 7-9\,10$^{-4}$ level, especially in 2014
\citep{alm16}.
This level of linear polarization would lead to a crosstalk with ESPaDOnS of about 2\,10$^{-6}$ , that is three times smaller than the 
signal observed on February 14, 2012, when the Stokes $V$ signal amplitude was about 6\,10$^{-5}$ (Fig.\,\ref{n1n2}). 
Therefore, this ESPaDOnS observation is effectively crosstalk-free.
Comparing with the Narval observation of \object{Betelgeuse} obtained a few nights earlier (i.e., well below the expected variation timescale, see Sect.\,5), 
on February 10, 2012, also presented in Fig.\,\ref{n1n2}, it is clear that both Stokes $V$ profiles have the same amplitude. 
This demonstrates that on February 10, 2012, the crosstalk of Narval did not significantly affect the Stokes $V$ signals.
Generally speaking, in the case of Narval, the crosstalk from the maximum observed linear polarization would be less than 2\,10$^{-5}$. 

\begin{figure}
\centering
\includegraphics[width=7 cm,angle=-90]{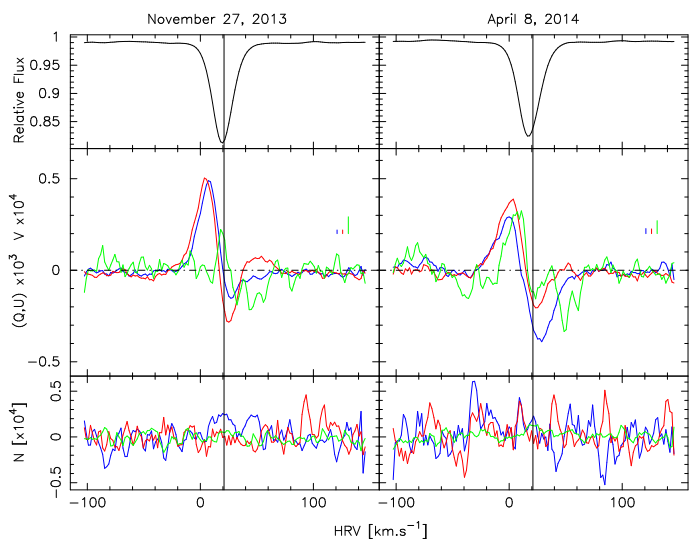} 
\caption{$IQUV$ LSD profiles of Betelgeuse for 27 November 2013 and April 8, 2014. 
Top: Stokes $I$. Middle: Stokes $Q$ (blue), $U$ (red), and $V$ (green) signals, 
where the $V$ signal has been magnified by a factor ten.
Bottom: null polarization signal $N_{1}$, with the same color code as for the Stokes parameters.
Small vertical lines represent typical error bars with the same color code.
The vertical line corresponds to the heliocentric radial velocity (HRV) of Betelgeuse (about 21\,\kms).}
\label{crosstalk}
\end{figure}

To generalize the weak influence of the crosstalk on the whole data set, we take advantage of our quasi-simultaneous Stokes $QUV$ 
Narval observations.
When both linear and circular polarization measurements were obtained within a two-day interval (corresponding to 17 sequences), 
we compared the extrema of the different signals.
It appears, from the analysis of these 17 $QUV$ sequences, that the $V$ signal amounts to between 4.5\,\% and 38\,\%
of the linear polarization, being 12\,\% on average - much larger than the expected crosstalk contribution.
Figure\,\ref{crosstalk} presents the measurements of $QUV$ signals for two nights, with the $V$ profile enhanced by a factor of ten with
respect to the $QU$ profiles.
A first remark is that the $N_1$ signal is flat for all Stokes parameters.
On November 27, 2013, the weak positive peak of Stokes $V$ at about 20\,\kms\ is not aligned with any $QU$ peaks (whatever their sign), 
and the aligned linear (absolute) value leads to $V/Q$ or $V/U$ of about 10\,\% - well above the 3\,\% crosstalk limit.
The situation is the same for the night of April 8, 2014, where all the peaks, aligned within a few \kms, show circular to linear ratios
again of the order of 10\,\%.
In addition, for both nights, the $V$ signal presents a structure located at a velocity of about 50\,\kms\ that is not present at all in the $QU$ signals
(possibly even extending outside the $I$ profile), and thus it cannot be attributed to a crosstalk effect from $QU$ to $V$.
Hence, we conclude that the detected Stokes $V$ signal is not due to, and only marginally affected by, crosstalk from the  
Stokes $Q$ and $U$ signals.

\subsection{Null polarization $N_{1,2}$ signatures associated with Stokes $V$ profiles of Betelgeuse}

In addition to the $QUV$ parameters, the LSD procedure provides two null polarization signals $N_1$ and $N_2$ 
\citep[see][]{dsc97}.
These profiles are used to diagnose spurious contributions to the polarization, in particular for 
weak Stokes $V$ profiles such as those of Betelgeuse. In principle they should be featureless, but we in fact observe nonzero signatures
in some of the $N$ spectra associated with our observations.
This is not uncommon, and can have a variety of explanations. For example, stellar variability or changing sky conditions during the four sub-exposures may induce a spurious signal in $N$. 
In addition, \citet{fpb16} found that for young cool stars, in the case of poor S/N, an nonzero $N$ signal could occur due to the very noisy 
blue part of the spectrum. 
This appears to be partly the case for our noisiest spectra, but even removing their bluest parts does not appreciably
clean the $N_{1,2}$ signals. 
\object{Betelgeuse} is not a rapidly varying star and we managed to observe it in good atmospheric conditions in order to detect the very weak Stokes $V$ signal ($\sim 1-2\,10^{-5}$). The detection of $N$ signatures is therefore something of a puzzle.

From the two close nights of February 2012 presented in Fig.\,\ref{n1n2}, it is shown that while the ESPaDOnS $N_{1,2}$ spectra are similar and unambiguously 
weaker than the $V$ signal, the corresponding Narval data present structures within both $N_{1,2}$ spectra that have amplitudes close to the circular 
polarization signal.
However, we note that both Stokes $V$ shapes are comparable for both considered nights, separated by less than four nights, that is, 
well below the expected timescale for variation of the polarimetric signal (see Sect.\,5).
Hence, since the Narval observation shows strong null polarization signals while the ESPaDOnS observation does not, we can infer that the $N$ behavior does not
significantly alter Narval's Stokes $V$ signal.

\begin{figure}[h]
\centering
\includegraphics[width=7cm,angle=-90]{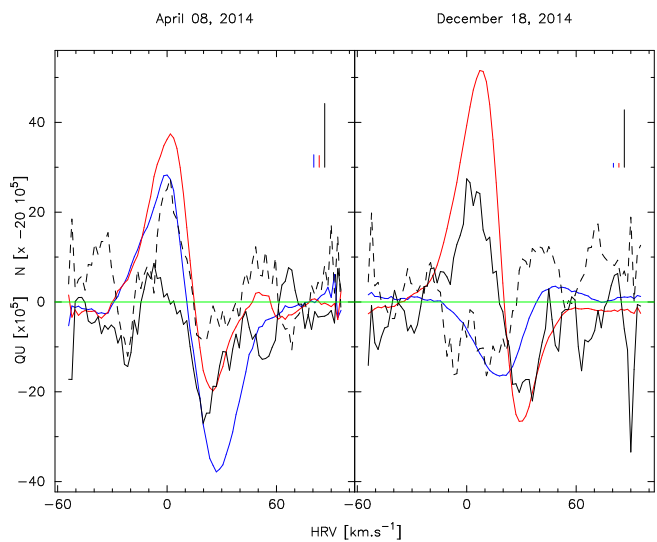}
\caption{Stokes $Q$ (blue), $U$ (red), and from the $V$ measurements $N_{1}$ (continuous) and $N_{2}$ (dashed) LSD profiles of \object{Betelgeuse} for 
April 08, 2014 (left) and December 18, 2014 (right) obtained with Narval.
Small vertical lines represent typical error bars with the same color code.
We note that the $N_{1,2}$ signals have been multiplied by a factor -20.}
\label{ctn1n2_2}
\end{figure}

The occurrence of significant $N_{1,2}$ signals in observations of \object{Betelgeuse}, both with Narval and ESPaDOnS 
\citep[and also observed by][]{gwh10} is more frequent than in any other star observed so far, apart from the RSG star
\object{$\mu$\,Cep} \citep{tlm17}.
These authors have shown that in the case of \object{$\mu$\,Cep}, when disentangling Zeeman and crosstalk contributions to Stokes $V$, 
the main part of the $N_{1,2}$ signals was contaminated by the linear polarization.
Looking closely at the observations of the magnetic Ap star $\gamma$\,Equ, which enables the disentangling of Zeeman and crosstalk contributions to 
Stokes $Q$ and $U$, we found the same effect: the $N_{1,2}$ profiles corresponding to both Stokes $Q$ and $U$ measurements can be obviously polluted by the strong 
Stokes $V$ signal at the $10^{-2}$ level.
As an example for Betelgeuse, Fig.\,\ref{ctn1n2_2} presents, for two nights for which both linear and circular measurements have been obtained, the null 
$N_{1,2}$ signals of the $V$ measurements together with the linear polarization $Q$ and $U$ signals. 
For both nights, at least one $N$ signal (multiplied by a factor of -20 for illustrative purposes) is very similar to the $U$ profile, and to a lesser
extent, to the $Q$ profile.
An interesting observation is that while $N_2$ mimics the $Q,U$ signals on the night April 8, 2014, it is the $N_1$ profile that is affected on the
night of December 18, 2014.
Therefore, it appears that the $N$ measurements are affected by the strong linear polarization in various ways. In particular, the $N$ profiles may
reflect the linear polarization by an amount estimated to be of about 2.5\,\%. 
Thus, structures present within the $N$ signals that are below 2.5\,\% of the $QU$ polarization amplitudes for a given wavelength may not be attributed 
to spurious extra-contributions.

\begin{figure*}[h]
\centering
\includegraphics[width=17cm,angle=0]{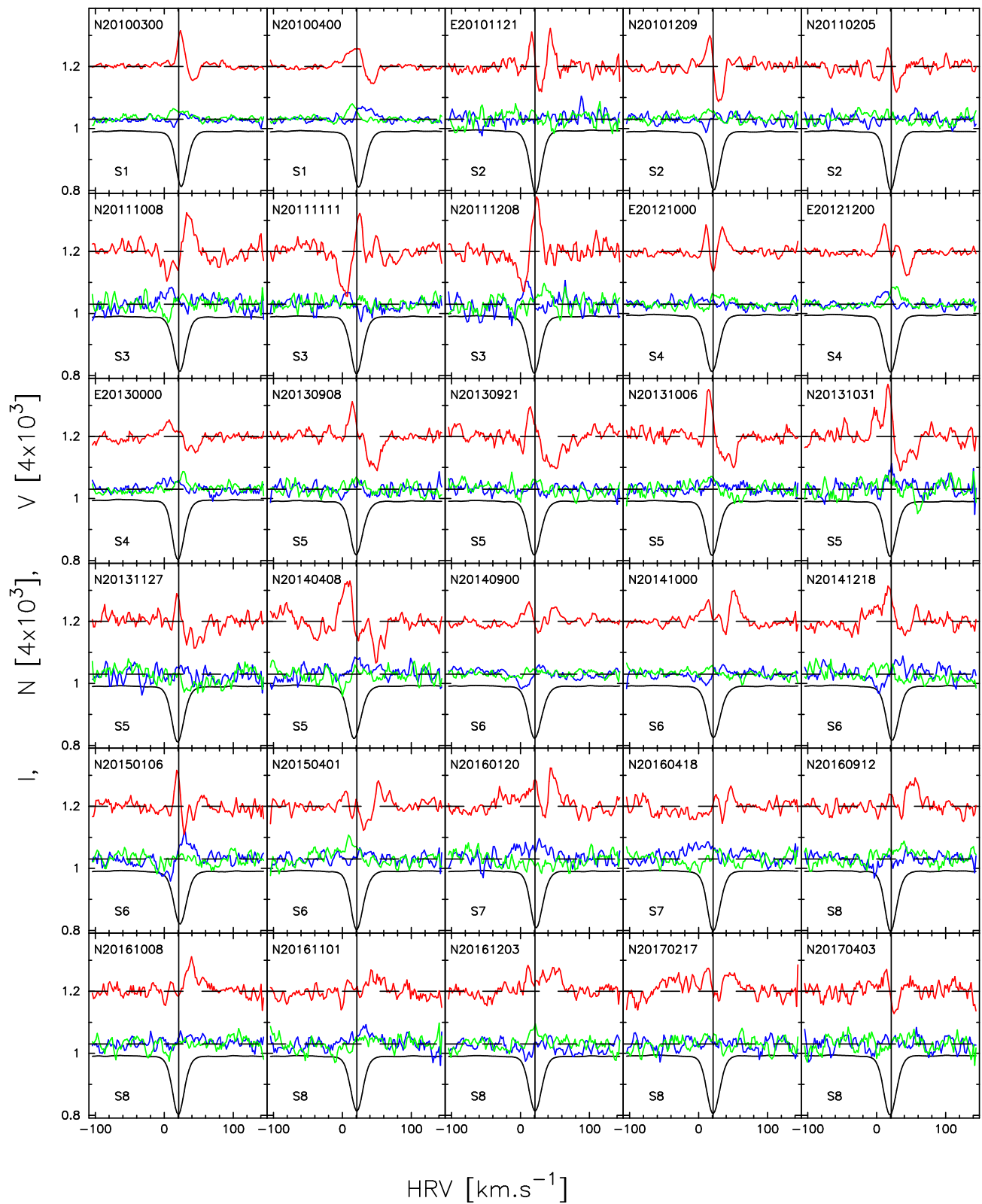}
\caption{Stokes $V$ (red), null polarization $N_{1}$ (blue) and $N_{2}$ (green), and $I$ (black) LSD profiles of \object{Betelgeuse} 
for the 30 dates selected as described in Sect. 3.2 and spanning along the eight seasons between March 2010 and April 2017 (Table\,\ref{tab1}).
We note that the UT observation date is encoded for each retained night (see Table\,\ref{tab1}).
The vertical line corresponds to the heliocentric radial velocity (HRV) of Betelgeuse (about 21\,\kms).}
\label{VN1N2}
\end{figure*}

Therefore, in order to restrict our analysis to what we presume to be the most reliable data, we decided to remove observations 
(i.e., the corresponding nights) that exhibit large and/or structured null polarization profiles.
This represents a relaxation of the criterion proposed by \citet{bll09} who suggested to reject all profiles with corresponding $N$ signal above the 3$\sigma$ level.
The retained data are provided in Table\,\ref{tab1} when the column ``label'' contains an entry.
Therefore, most of the data in Table\,\ref{tab1} will be used for the work on Stokes $I$ profiles (Mathias et al., in preparation), but our results on 
Stokes $V$ profiles will only be based on spectra with $N_{1,2}$ profiles presenting signals significantly weaker than the corresponding circular polarization 
signals, which represents 30 epochs. 
The 30 LSD profiles for Stokes $I, V$ and $N_{1,2}$ are illustrated in Fig.\,\ref{VN1N2}.
We note that a three-pixel filtering is performed on the profiles to enhance visibility of the signal with respect to the noise.  


\section{Properties of the Zeeman signal}

\subsection{Evaluating the Zeeman origin of Stokes $V$}

Since magnetic fields have already been detected in a large number of cool and evolved stars through spectropolarimetric observations
(see Sect.\,1), it is very likely that the circular polarization detected in \object{Betelgeuse} 
is due to the Zeeman effect as well. 
The general shape of the Stokes $V$ signal (e.g., in observation N20111208) is actually reminiscent of that predicted by the Zeeman effect \citep{m93} 
in the case of a single spot with homogeneous velocity and magnetic field.
However, since the level of the signal is very weak, following \citet{bpl16}, we performed different tests to ascertain the Zeeman nature 
of the Stokes $V$ signal, by comparing different LSD profiles computed from selected masks.
Indeed, in the first order approximation, valid for weak fields and safely applicable to most cool stars, Stokes $V$ is proportional to the
derivative of the intensity with respect to the velocity $dI/dv$ (i.e., the depth), to the wavelength, and
to the effective Land\'e factor $g_{\rm eff}$. 
We note that this approximation may be insufficient in some cases, for instance for profiles distorted by a complex velocity field.

\begin{figure}[h]
\centering
\includegraphics[width=7 cm,angle=0] {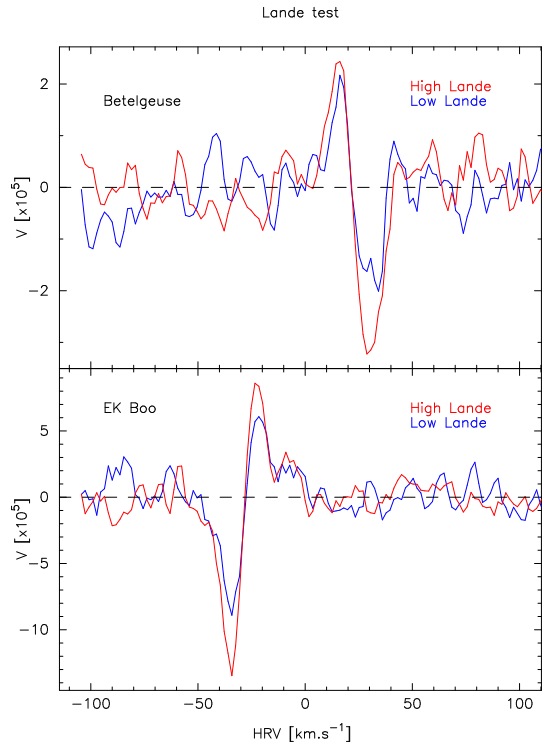} 
\caption{Comparison of Stokes $V$ profiles obtained by selecting photospheric lines of low (blue line) and high (red line) magnetic sensitivity
according to the Land\'e factor. 
Top: Betelgeuse on December 9, 2010. Bottom: The magnetic M giant EK\,Boo on March 18, 2009.}
\label{zeemantest}
\end{figure}

For a star with negligible rotational broadening of its spectral lines, the ``classical'' Zeeman picture predicts that lobes of positive and negative signs should be observed, resulting in the well-known characteristic $S$ shape of the $V$ signal.
A few nights that present such an unambiguous $S$ shape were selected, and we performed tests in particular according to the
proportionality of the $V$ signal with the Land\'e factor, which quantifies the magnetic sensitivity.
The complete mask described in Sect.\,2 was split into two sub-masks (about 6\,000 lines each) of low and high Land\'e factors, the separation 
being at $g_{\rm eff}=1.207$ as an average of the whole mask.
The resulting LSD profile was then normalized in order to obtain the same Stokes $I$ profile depth.
Since the Stokes $V$ signal is weak in the case of \object{Betelgeuse}, the tests were compared to an M giant star for which a magnetic 
field (of about 5\,G) has been confidently detected, \object{EK\,Boo} \citep{kac10}. 
Figure\,\ref{zeemantest} displays a representative example of this comparison.
As expected for a Zeeman effect, we note a clear increase of the Stokes $V$ signal for the high-Land\'e sub-mask ($<g_{\rm eff}>=1.425$) 
compared to the low-Land\'e mask ($<g_{\rm eff}>=0.878$).

Such tests have also been performed for the two other parameters: line depth and wavelength. 
In the first case, the sub-masks also confirm the Zeeman effect, but the increase of the signal with line depth is not a unique characteristic 
of Zeeman effect, as it is also observed in the case of scattering processes \citep{alm16}.
As for the second case, because of the low temperature of the star, LSD profiles computed from the blue sub-mask were very noisy and did not 
lead to convincing results. 

\subsection{Zeeman polarity and crossover profiles}

Interpreting our Stokes $V$ signals as due to Zeeman effect in the case of weak magnetic fields enables us to study their $S$ shape 
\citep{m89,m93} and to infer their polarity. 
In this work, the positive Stokes $V$ polarity corresponds to a positive first (blue) lobe; the negative Stokes $V$ polarity corresponds to a negative 
first (blue) lobe. 
We see in Fig.\,\ref{VN1N2} that the polarity changes between positive (e.g., N20100300, N20130908) and negative (e.g., N20111008).
In addition, configurations similar to crossover effect \citep[composition of two opposite polarity signals, ][]{b51} also occur
(e.g., E20101121, N20110205, E20121000). 
Furthermore, the peculiar shape of the detected crossovers is observed at the required time of the polarity change: the Stokes $V$ profile is symmetrical with 
respect to the radial velocity inferred from the $I$ profile.
However, if the Stokes $V$ profile shape may be more or less interpreted in terms of a classical Zeeman signal until 2014, the situation becomes unclear after
that date as many $V$ signatures (e.g., N20160912, N20161203) present nonstandard structures.

\begin{figure}[h]
\centering
\includegraphics[width=7 cm,angle=-90]{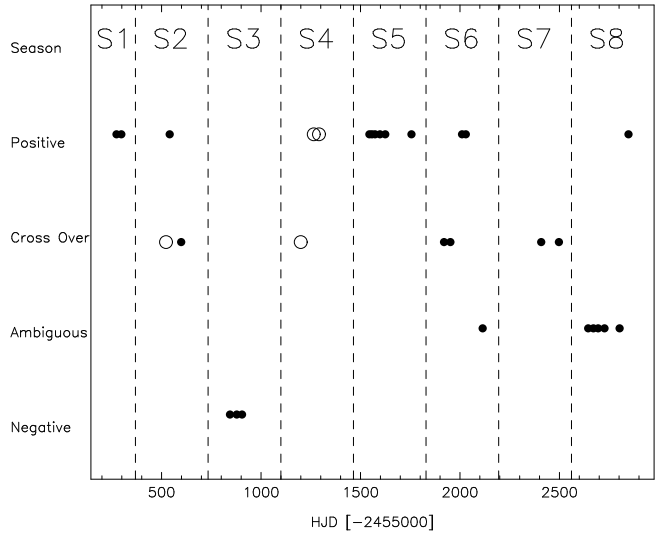} 
\caption{Variations of polarity during the eight observation seasons 2010-2017 (S1 -- S8). 
Open and full symbols represent ESPaDOnS and Narval, respectively.}
\label{Vlong}
\end{figure}

A summary of the interpretation of the $V$ measurements is represented in Fig.\,\ref{Vlong}.
The polarization has changed at least three times between March 2010 and April 2014.  
During these five seasons the Stokes $V$ signal mainly presents two lobes, as in the classical Zeeman shape. 
The crossover profile itself has occurred three times during these five observing seasons, twice in the 2010/2011 season (Season 2) and once in 2012 (Season 4).
During the three next seasons (Seasons 6 to 8) the shape of the Stokes $V$ profile is more ambiguous: except for December 2014 and January 2015 when a positive 
polarity clearly appears, the shape of the Stokes $V$ profile is much more complex.

\subsection{The longitudinal magnetic field $B_\ell$}

\begin{figure}[h]
\centering
\includegraphics[width=7 cm,angle=-90]{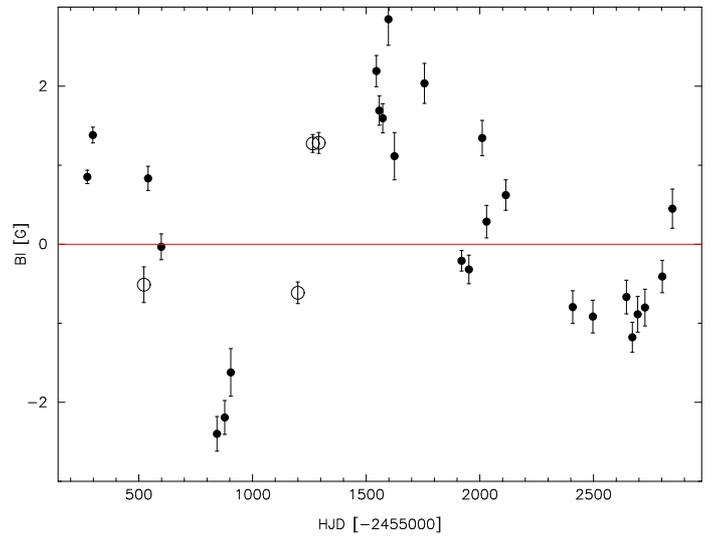} 
\caption{Evolution of the longitudinal component $B_{\ell}$ during the 8 seasons. 
Open and full symbols represent ESPaDOnS and Narval, respectively.
}
\label{Blong}
\end{figure}

The surface-averaged longitudinal magnetic field $B_\ell$ may be computed from the $V$ signal using the first-order moment
method  \citep{rs79}, adapted to LSD profiles \citep{dsc97,wdl00}.
However, this method assumes that Stokes $I$ and $V$ share a common center of gravity, which is usually not the case for Betelgeuse, since the polarization
signal is found to be redshifted \citep[][, see also Sect.\,6]{pak13}.
In addition, the $I$ signal is asymmetric, its shape being modified in particular by the convective velocity field \citep[Mathias et al., in preparation]{jp07}.
Finally, and especially during Seasons 6-8, the shape of Stokes $V$ is very complex and cannot be assimilated to a profile having 
unambiguous parameters on the stellar surface, meaning one not characterized through a single location and/or field intensity.
Therefore, the ``classical'' (i.e., first-order moment) measurements of $B_{\ell}$ can only provide an estimate of the magnitude of an averaged longitudinal 
magnetic field, having a value around 1\,G. 
The $B_{\ell}$ measurements obtained using the first-order moment method are illustrated in Fig.\,\ref{Blong}.
This complements the work presented by \citet[their Fig.\,1]{bpa13}. 
We underscore the fact that the interpretation of these measurements aisre not as straightforward as in stars having a localized magnetic field such as Ap stars, see
Sect.\,6 \& Sect.\,7.

\section{Timescales of variation of the Stokes parameters}

Red supergiants often present semi-regular variations that have been attributed to radial stellar pulsations  \citep[e.g.,][]{s72} and/or
to largescale convection in the envelope \citep[e.g.,][]{s10}.
For Betelgeuse, many attempts to derive periods have been carried out; the most extensive of which is that of \citet{ksb06} who analyzed
an extended data set from the AAVSO photometric database.
This latter study shows that, as other RSGs, \object{Betelgeuse} presents two variation timescales: a relatively fast one, of the order of 400\,d, 
that can be attributed to a fundamental or low-order overtone radial mode, and a relatively slow one, of the order of 2000\,d, often referred 
to as the LSP.
These LSPs, well known in the less-massive AGB stars, are still poorly understood. 
Common interpretations involve binarity, nonradial $g$-modes or magnetic activity \citep{wok04}.

From Fig.\,\ref{Vlong}, we see that polarity changes occur with timescales ranging from about 3 months 
(between the two crossover phases observed in 2010) to about two years, between October 2012 and September 2014 (hypothesing that there is no
rapid change of polarity between the observing seasons). 
While these changes appear very clearly during Seasons 1--5, the polarity of the Stokes $V$ parameter is more challenging to determine for Seasons 6--8.
It may be thus interesting to investigate if a polarity change timescale is present or not, and if this timescale can help to understand the
origin of the $V$ signal variations.
So, despite our relatively short dataset for the Stokes $V$ measurements (about 7.5\,years), we attempted a frequency analysis.

\begin{figure}[h]
\centering
\includegraphics[width=7 cm,angle=-90]{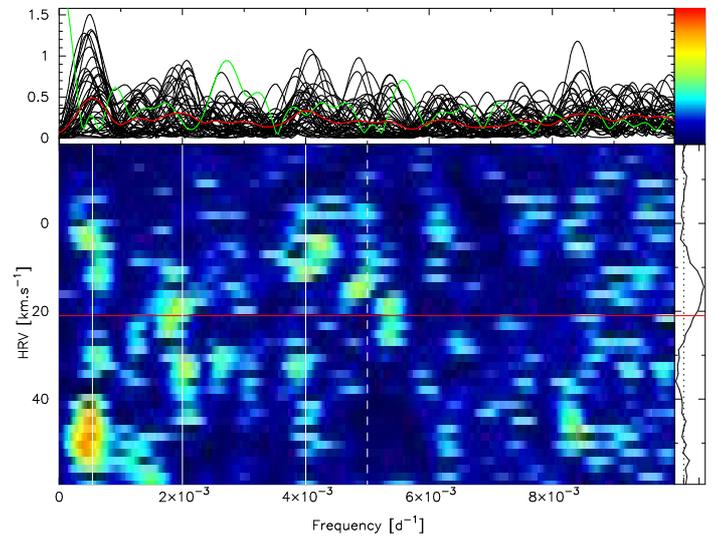}
\caption{2D Fourier analysis of the Stokes-$V$ parameter. 
The horizontal red line represents the star velocity, 21\,km.s$^{-1}$.
The vertical white bars represent the three frequencies around 1850\,d, 500\,d and 250\,d, while the dotted white line
marks a signal around 200\,d.
The above picture shows, for each velocity-bin, the Fourier periodograms together with its average (red) 
and window function (green).
On the right is shown the mean-$V$ profile together with the star velocity (red), while the dotted line 
represents the null polarization level.}
\label{2DF_Vret}
\end{figure}

We thus considered the data set corresponding to the 30 retained nights (see Table\,\ref{tab1}) and we proceeded to apply a standard 2D Fourier 
analysis \citep[CLEAN,][]{rld87} up to a frequency $f=0.01$\,d$^{-1}$.
Results are presented in Fig.\,\ref{2DF_Vret}.
Signal is present in three main regions: one at a very low frequency (about $5.4\,10^{-4}$\,d$^{-1}$, or 1850\,d), and two at about 
$2\,10^{-3}$\,d$^{-1}$ (500\,d) and $4\,10^{-3}$\,d$^{-1}$ (250\,d), this latter being a harmonic. 
Considering the uncertainties on the periods are 900\,d, 100\,d and 30\,d, the low frequency is in agreement with the 
LSP, while the 500\,d period could be related to the ``fast'' variation scale around 400\,d.
It is interesting to note that the low-frequency signal occurs both inside and outside the $V$ profile, in particular around a 
velocity of 50\,\kms\, as already noted in Sect.\,3.1.
Thus, this highly redshifted $V$ signal is not likely to be noise.
There is additional signal located clearly within the profile around $5\,10^{-3}$\,d$^{-1}$ (200\,d), but the two close frequencies involve
different parts of the $V$ profile.
These latter frequencies, as well as other structures (e.g., around $8.5\,10^{-3}$\,d$^{-1}$ or 120\,d) may be real, but are located too close to the peaks 
of the window function to be firmly established.
We note that while any Fourier analysis assumes the variations to behave as sinusoids, a further analysis performed with the PDM method \citep{s78} 
points toward similar frequencies.
Therefore, both photometric periods already indentified for Betelgeuse \citep{ksb06} seem to be present in the Stokes $V$ profile variations, 
in addition to a shorter one around 250\,d.

\begin{figure}[h]
\centering
\includegraphics[width=7 cm,angle=-90]{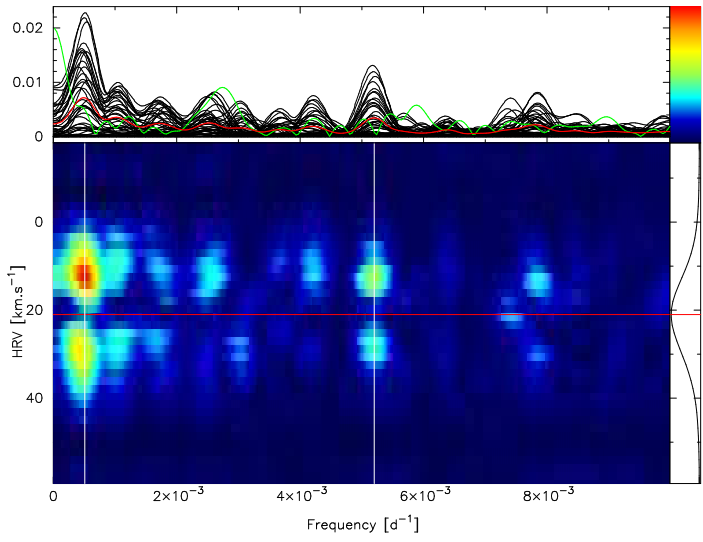} 
\caption{2D Fourier analysis of the Stokes-$I$ profiles. 
Legends are as for Fig.\,\ref{2DF_Vret}.}
\label{2DF_LSD}
\end{figure}

The Stokes $I$ profiles have a much higher S/N (more than 40\,000) and are much more numerous 
than any other Stokes parameter.
We therefore followed the same frequency analysis as for Stokes $V$; the results are presented in Fig.\,\ref{2DF_LSD}.
Again, a prominent peak, well inside the $I$ profile, is present at low frequency ($5.1\,10^{-4}$\,d$^{-1}$, 1960\,d), but also at 
twice this frequency, probably as a harmonic.
Another important peak is present at $5.2\,10^{-3}$\,d$^{-1}$ (200\,d). This frequency is also present in Fig.\,\ref{2DF_Vret}, though close to a 
peak of the window function.
Conversely, the peaks at $2\,10^{-3}$\,d$^{-1}$ and $4\,10^{-3}$\,d$^{-1}$ do not appear in Fig.\,\ref{2DF_LSD}.
We also note that these variations appear essentially within the line wings, suggesting an important 
tangential component (i.e., orthogonal to the stellar radius) of the velocity field, probably due to convection.
However, other mechanisms may also be in action, such as changes in line profile due to molecular veiling.

Although the extent of the Stokes $Q$ and $U$ measurements is relatively short (less than 3.5\,a), we undertook a rough frequency analysis.
For both parameters, there is abundant power at the low frequency detected in the $I$ and $V$ profiles.
The peak around $5\,10^{-3}$\,d$^{-1}$ (200\,d) is present in the $Q$ variations, but totally absent in the $U$ variations.
A signal is also present around $3\,10^{-3}$\,d$^{-1}$ (350\,d) in both Stokes $Q$ and $U$.
Finally, a similar analysis of the total linear polarization modulus $\sqrt{Q^2+U^2}$ confirms the $5.4\,10^{-4}$\,d$^{-1}$ frequency (1890\,d)
as the dominant peak.

From this crude frequency analysis, it appears that the usual ``short'' 400\,d period is hardly detected within the different Stokes parameter
profiles; a signal is present between 120\,d and 500\,d.
An obvious explanation is that the window function is quite prominent around this frequency, and prevents any clear detection in this frequency region.

The situation is clear for the low frequency, which corresponds to the 2000\,d photometric LSP.
\citet{ksb06} noted that a strong $1/f$ noise component is present in the power spectra of the brightness fluctuations of RSGs, that might lead to
the development of random peaks at the lowest frequencies.
However, the fact that a distinct peak exists, common to all Stokes parameters which have a different sampling, suggests the reality of this
LSP in spectropolarimetric measurements.
In addition, this period being common to both linear and circular polarization suggests a link between the two phenomena that are expected to have
different physical origins, respectively scattering and magnetism.
However, we note here that the length of our timesseries prevents a detailed, high-resolution frequency analysis, and additional continuous 
observations over the coming years will be important to refine the different timescales discussed in this paper.

\section{The spot model}

A star with a localized monopolar magnetic spot on its surface is expected to show the classical $S$-shaped Zeeman signature \citep{m93}.
In particular, the right and left circular polarization components should be anti-symmetric with respect to the $I$ line profile.
As presented in Fig.\,\ref{Vlong}, the five first seasons of observation of \object{Betelgeuse} present one dominant polarity and a shape compatible with that expected from
the presence of a magnetic spot.
However, there are two facts that weaken this simple view, as illustrated for example, by observations N20100300 or N20130908 (Fig.\,\ref{VN1N2}).
First, the Stokes $V$ profiles are not centered on the rest frame velocity, but are often redshifted by about 10\,\kms 
(e.g., Fig.\,\ref{2DF_Vret}), indicating that the magnetic field is located in particular regions of the star that favour downward flows.
Second, the left and right circularly polarized fluxes appear to be unequal in absolute value.
These observed properties could be interpreted as implying the presence of additional magnetic locations, characterized by different Zeeman intensities
and different velocities.
Indeed, a configuration with two regions of opposite polarity is commonly seen in Ap stars, even with weak longitudinal magnetic fields and low projected
rotation velocity \citep{aws07}, and leads to crossover structures, such as that observed in observation E20121000 (Fig.\,\ref{VN1N2}).
However, both the redshifted and asymmetric shape of the Stokes $V$ profiles also applies for most of the observed ``crossover'' configurations (e.g., 
E20101121 or N20160120), which are unsatisfactorily modeled with one or two equal ``spots''.
Moreover, the behavior of the Stokes $V$ profiles is not easily interpreted in Seasons 6--8 (Fig.\,\ref{Vlong}), showing very complex structures,
definitely far from classical $S$ or crossover shapes.
A natural interpretation would be to consider a mix of different magnetic field concentrations having different polarities and velocities.
The fact that the signal is weakened by roughly a factor of two for the two later seasons could be due to a dilution of the ``spot(s)'' of the first seasons.
We emphasize that during this same period, the linear polarization signal in both Stokes $Q$ and $U$ is also weakened by a similar factor, although the
profiles shapes remain roughly unchanged \citep{alm16}.
This strengthens the link between circular and linear polarizations pointed out in the previous section.
\citet{alm16} interpreted the linear polarization variations as due to two main bright spots on Betelgeuse.
In the following, we continue to apply this spot model to the newly-obtained linear polarization observations described in Table\,\ref{tab2}, and tentatively
extend it to the circular polarization observations.

\subsection{Linear polarization}

From the shape of both $Q$ and $U$ signals, \citet{alm16} were able to reconstruct the location of the scattering centers at the origin of the linear
polarization, leading to two bright spots both near the eastern and southern limbs of the stellar disk (with a $180^\circ$ uncertainty).
This model was fully compatible with quasi-simultaneous high angular resolution observations with VLTI/PIONIER that detected the emergence
and variation of a large hot spot at the eastern limb of \object{Betelgeuse} \citep{m14,mkp16,okh17}. 

\begin{figure*}[h]
\centering
\includegraphics[width=\textwidth,angle=0.]{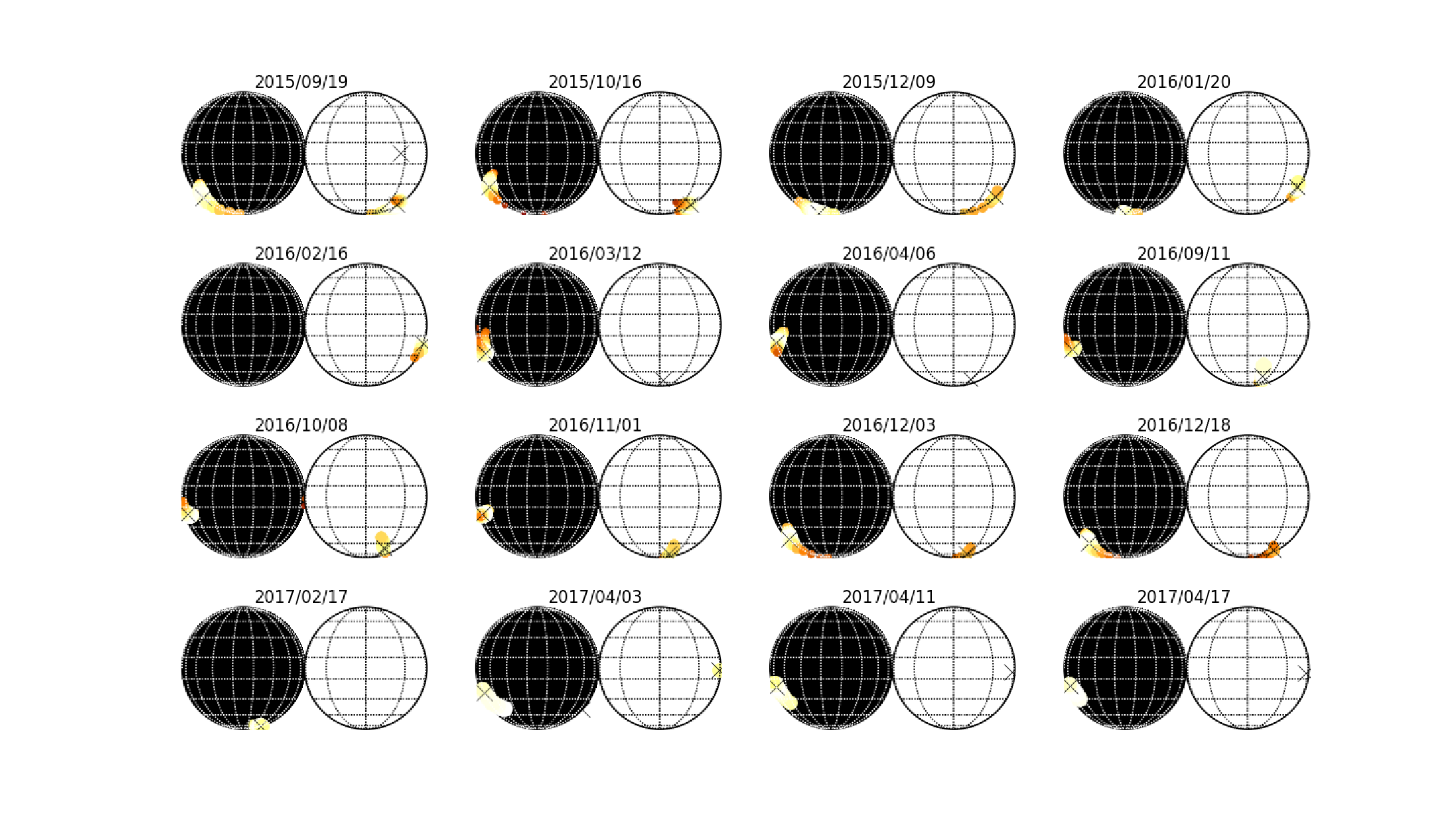}
\caption{Sequence of images of Betelgeuse from linear polarization for the dates given in Table\,\ref{tab2}. 
North is up and east is left for all images.
For each date the image (relative intensities) on the left represents the blueshifted signals (spot1) which are located on the visible hemisphere; 
the image on the right represents the redshifted signals (spot2) which are supposed to correspond to the opposite hemisphere. 
The crosses show the positions corresponding to the maxima of the linear polarization. 
The model used is described in \citet{alm16}.}
\label{maplin}
\end{figure*}

Using our new linear polarimetric data, we used the same model \citep{alm16} to follow two spots until April 2017.
The angular parameters derived from the model are provided in Table\,\ref{tab2}, and resulting maps are presented in Fig.\,\ref{maplin}.
Even though the signal has decreased in intensity by a factor of about 2.5, the areas of the scattering centers are still more or less located near
the eastern and southern limbs on the stellar disk, reflecting a long-term stability of the bright spots, compatible with the results of Sect.\,5.

\subsection{Circular polarization}

\begin{figure}[h]
\centering
\includegraphics[width=10.cm,angle=0.]{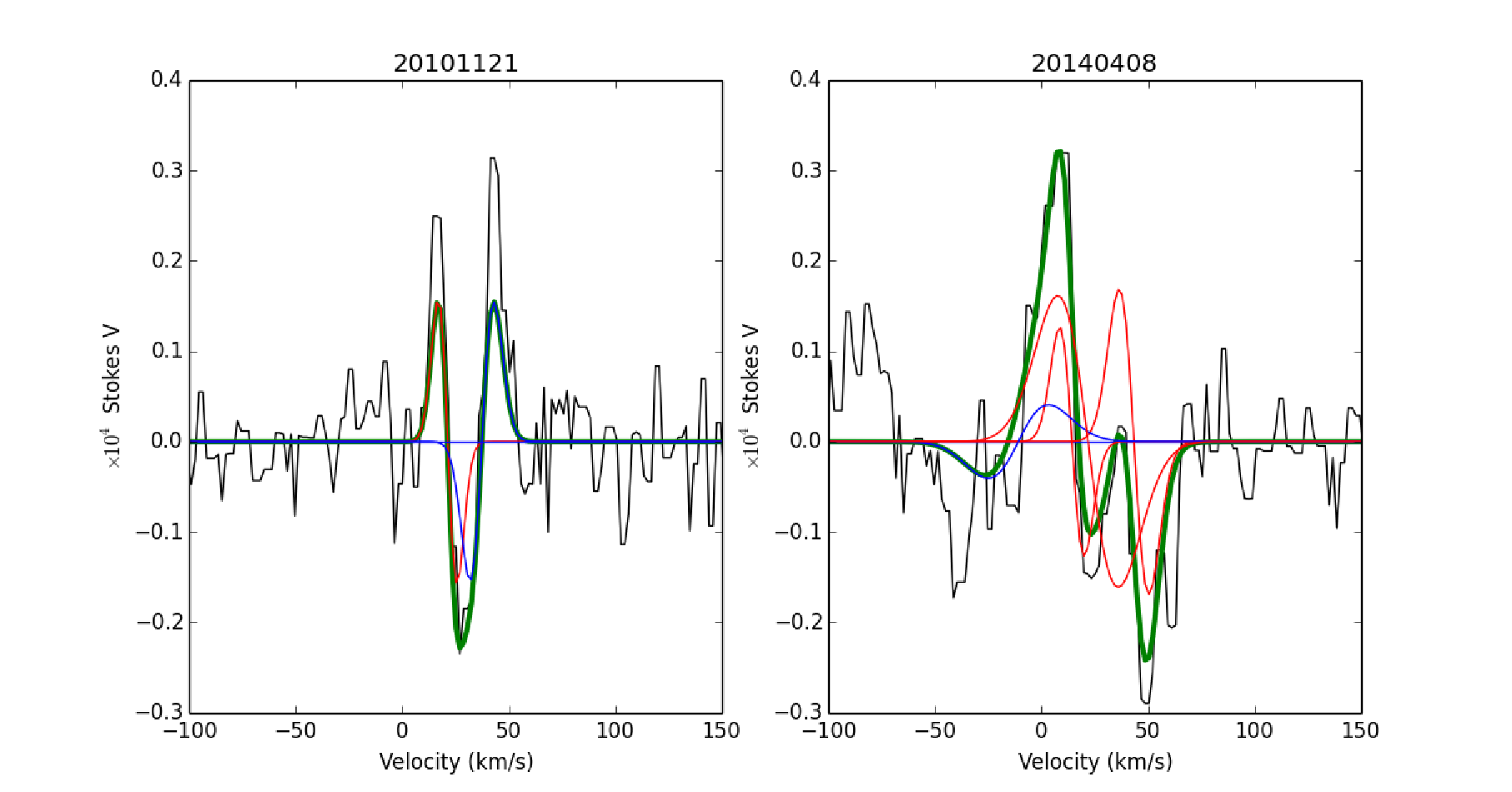}
\caption{Fit of the $V$ profiles for the nights E20101121 and N20140408.
Both positive (red) and negative (blue) magnetic components of each considered magnetic spot, while the total fit is
represented with the green line.
Whereas the best model for E20101121 requires two magnetic spots of opposite polarities, N20140408 is represented through four magnetic spots.}
\label{fitV2}
\end{figure}

\begin{figure*}[h]
\centering
\includegraphics[width=\textwidth,angle=0.]{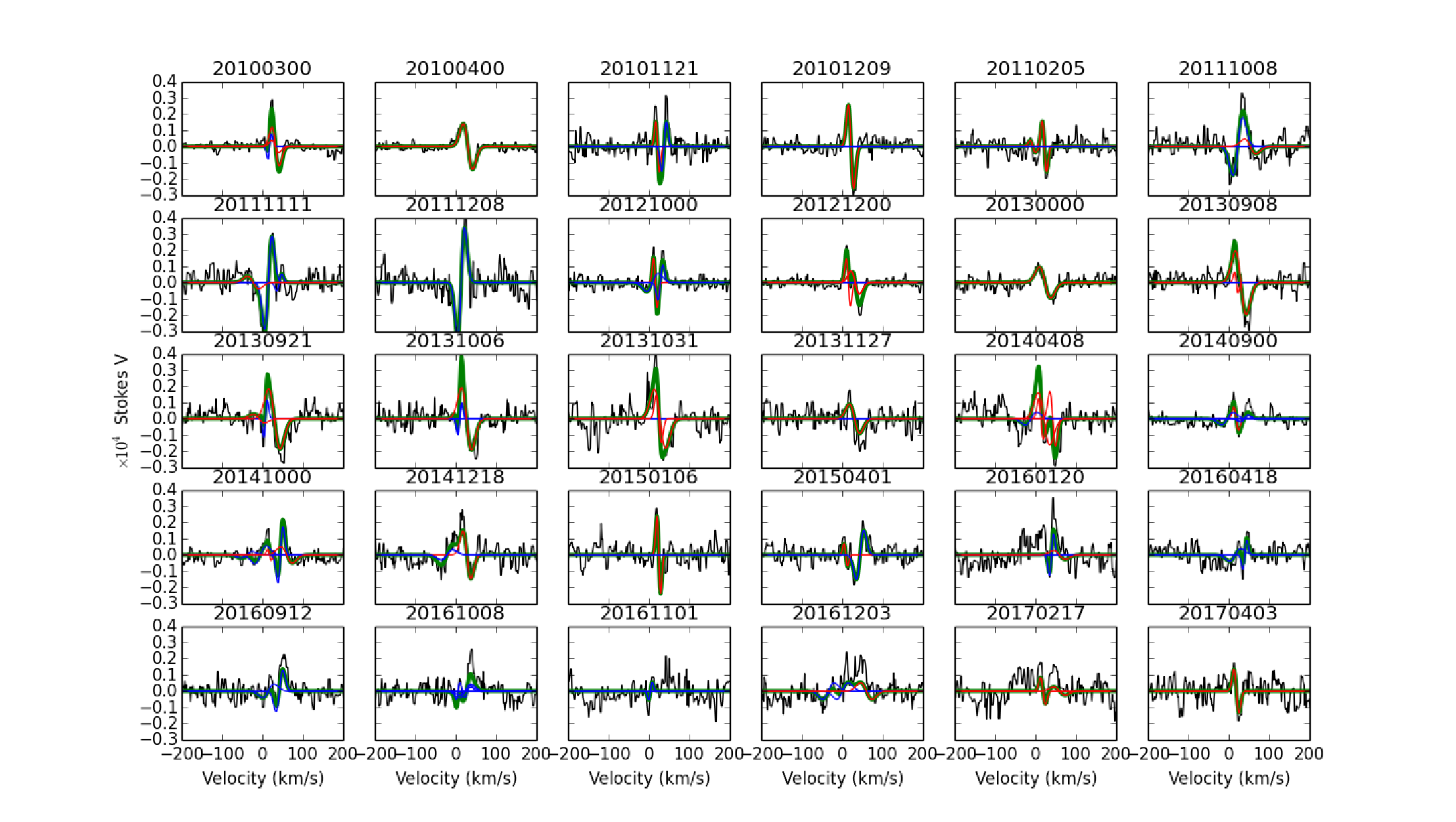}
\caption{Fit of the $V$ profiles for the observation dates represented in Fig.\,\ref{VN1N2}.
Both positive (red) and negative (blue) magnetic components are represented for each contributing spots in a given night, 
while the sum of these contributions is the green line.}
\label{mapV}
\end{figure*}

Despite the complex magnetic structure expected of a star such as Betelgeuse, rough comparisons performed with solar magnetograms, in addition to
magnetohydrodynamics simulations both for main sequence stars \citep{bsc15} or directly related to Betelgeuse \citep{d04}, show that while complex, the 
magnetic structures may appear concentrated.
Whereas linear polarization is related to scattering anisotropy, interpreted as bright spots, circular polarization can be modeled using
several magnetic spots on the stellar surface, mimicking such magnetic concentrations.
Each magnetic spot is here characterized using three parameters. 
First, the width (FWHM) of the left-or-right polarization signal, that should be of the order of that of the Stokes $I$ profile that is, about 20\,\kms.
Second, the amplitude ($A$) of the magnetic field should be adjusted through the extrema of the Stokes $V$ profile.
These two parameters lead to the classical $S$ shape, that could be Doppler shifted, the associated velocity (HRV) representing the
radial motion of the considered spot and being thus the third parameter.
Therefore, each spot is modeled using the (FWHM, $A$, HRV) parameters.
Of course, the complexity of the $V$ signal requires more than one spot most of the time.
This is illustrated in Fig.\,\ref{fitV2}, where the resulting fits are presented for two nights (E20101121 and N20140408).
For night E20101121, the $V$ profile looks like a classic crossover and as expected two spots of opposite polarity (blue and red profiles) 
provide an acceptable fit (green profile) superimposed on the observed one.
It should be emphasized that each fit has been constrained to the minimum possible number of magnetic spots, through Bayesian marginalization, 
thus minimizing the risk of overfitting and/or to derive a nonunique solution.
The fit associated to night N20140408 requires the presence of four spots, three of positive polarities (red) and one of negative polarity (blue).
The fitting procedure was applied to the 30 selected nights presented in Fig.\,\ref{VN1N2}, and the results are illustrated in 
Fig.\,\ref{mapV}.
Compared to the linear polarization Stokes parameters, the Stokes $V$ signatures appear more complex, requiring from one (e.g., N20111208) to five
(e.g., N20160120) magnetic spots.
However, due to both the weak signal and the simplicity of the model, many configurations are doubtful or even obviously wrong,
such as the case of for example, N20140408 or N20161203. 
Indeed, considering for instance N20140408, the positive polarity does not correspond to any signal.

\begin{figure}[h]
\centering
\includegraphics[width=8.cm,angle=0.]{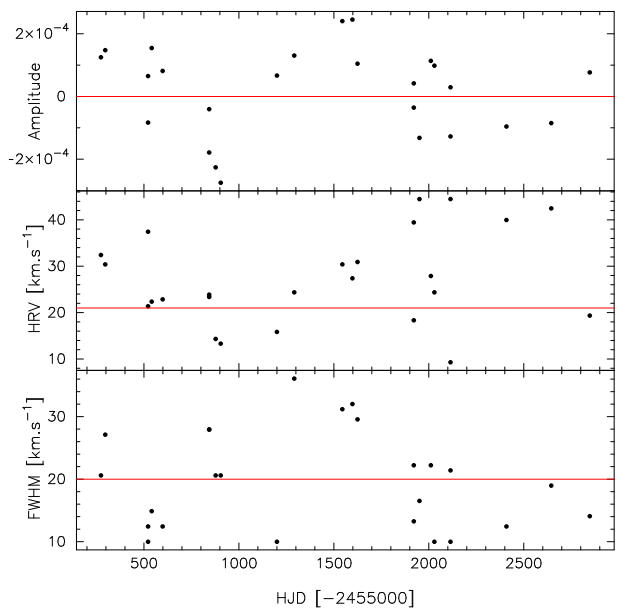}
\caption{Amplitude, Doppler velocity, and FWHM of each magnetic spot considered (one or two spots per retained night, see text). 
The stellar rest frame velocity and the mean Stokes $I$ FWHM are indicated by an horizontal red line.}
\label{fitV}
\end{figure}

We then retained the nights where one or two significant spots (i.e., having amplitudes significantly larger than was typical on 
a given night), with realistic widths compared to that of the $I$ profile, and we studied the three fitting parameters (FWHM, $A$, HRV) 
associated with each spot.
On this basis, nights E20140408, E20140900, E20141000, and all nights of 2016 were excluded.
The variations derived from the fitting parameters are illustrated in Fig.\,\ref{fitV}.
The first result is that the amplitude $A$ (the dominant polarity) clearly varies on the LSP timescale 
(about 2000\,d).
The velocities associated with the magnetic spots have variations closely linked to the amplitude variations, and appear redshifted most of the time 
(the average spot velocity is about 30\,\kms\ i.e., 10\,\kms\ above the radial velocity of Betelgeuse), 
as already noted by \citet{pak13}.
We note that this velocity corresponds to the center of the $S$ shape, and therefore each lobe extends to $\pm 20$\,\kms\ and explains the signal
already pointed out around 50\,\kms (see Sect.\,3.1 and 5).
Thus, conversely to the expansion velocity used by \citet{alm16}, we have to deal with structures that appear to be associated with downflows 
with a mean velocity of about 10\,\kms.
Finally, the width parameter appears quite dispersed, spread between 10 and 35\,\kms, whereas the Stokes $I$ profile remains within a 2\,\kms\ interval
around the mean value of about 20\,\kms, as already noted during the comparison of the signal location between Figs.\,\ref{2DF_Vret} and \ref{2DF_LSD}.
In contrast with the width of the $I$ profile, which is integrated over the whole visible disk and is thus relatively stable, the spectral line parameters 
associated with the left and right polarization for a given magnetic spot appear more localized and hence more sensitive to the local temperature,
turbulence or velocity field.

\section{Discussion and conclusions}

Our observations and modeling establish the presence of a magnetic signal that varies with timescales similar to those of the linear
polarization measurements, typically around 2000\,d and 300\,d.
All these polarization signals have been interpreted here in terms of spots, either magnetic spots for the circular polarization or bright spots for 
the linear polarization.

The complex pattern of the Stokes $V$ signal might be explained by several magnetic spots, each presenting a classic Zeeman
profile ($S$ shape) and shifted relative to one another by a given velocity.
In Ap stars, the velocity field associated with the Stokes $V$ variations, and in particular the crossover, is entirely dominated by the rotation of the star.
For \object{Betelgeuse}, the crossover morphology evolves on a timescale of a few months, much less than the expected rotation period of the star, 
around 17\,a \citep{udg98} or even 31\,a \citep{kdr18}, so that we can exclude such a scenario. 
The velocity field should thus concern (sub-)photospheric motions, such as pulsation and/or convection.
For cool evolved stars, the favoured pulsation modes are radial, meaning that the whole surface of the star expands or contracts on a timescale of typically a few
hundred days, at velocities of a few \kms\ \citep[e.g.,][]{wok04}.
We note that nonradial modes are not excluded, but they should be associated with very low-amplitude motions and would presumably be undetectable.
Despite our rough frequency analysis, it appears that the main timescale detected in the polarization parameters is much longer than the radial fundamental 
period (about 400\,d), and hence is rather linked to the Long Secondary Period (LSP).
This LSP timescale, about 2000\,d for RSGs, may be associated with giant convection cells, that might extend to typically the stellar radius, and have
lifetimes of the order of years \citep{cpj09,chy10,cpj11b}.

Giant convection cells as an explanation for the LSP have been studied by \citet{s10}, who provided the typical turnover timescale and velocity
of about 2500\,d and 7\,\kms\ for Betelgeuse, respectively.
\citet{alm16} explained the linear polarization through scattering anisotropies due to bright spots. 
In order to locate these spots on the surface, they introduced an arbitrary expansion velocity, that might be due either to the ascending flow within
the bright spot or within the atmosphere, propelled by the locally enhanced luminosity.
The circular polarization, being most of the time redshifted, cannot be associated with such an upflow at the center of the bright spot, but it
may be associated to the downflow at the edge of the same giant convection cell.
The ``concentration'' of magnetic field at the border of the cell may be the result of the advection flows that act on the magnetic regions  
and thus on their locations.
Such a phenomenon has been studied in the case of the solar supergranulation \citep{rmr16} in order to explain observations showing that a significant part 
of the flux that appears inside supergranular cells is observed to move toward the photospheric network at their boundary \citep{gbo14}.
Apart from the Sun, magnetohydrodynamic simulations of the upper layers of convective envelopes of cool main sequence stars have shown that the strong 
concentration of the magnetic flux in some of the convection downflows leads to a local increase of the field strengths by a factor of 100 \citep{bsc15}.
In particular, these authors show that the velocity field is responsible for the very inhomogeneous structure of the magnetic field: 
horizontal outflows from the granules keep the magnetic flux in the downflow regions, while the upflows become nearly field-free.
In addition, detailed numerical simulations show that, for Betelgeuse, the field becomes concentrated into elongated
structures much thinner than the scale of the giant convection cells \citep{df03}.
Thus solar observations and simulations of cool stars provide support for the hypothesis that the magnetic field is located at the edge (i.e., in the sinking part)
of the giant convection cells.
Our simple model reduces this distribution of magnetic concentrations to just two or three ``spots'' that suffice to reproduce the observations but 
that should be interpreted as a representation of a more complex distribution.
The spot velocity is sometimes blueshifted; this could be due to a perturbation of the downflow velocity, the motion of a neighboring cell,
or the particular projection of the radial velocity, for instance superposed to the rotation velocity, which can amount to 15\,\kms\ following \citet{udg98}.
We note that the velocity field of RSGs might be very complex, as already observed for Betelgeuse in the near IR \citep{owm11}, and also as in the case 
of Antares that presents strong upward and downward velocities across its photosphere and/or its (extended) atmosphere \citep{owh17}.
In addition to the direct link with giant convection cells, both the circular and linear polarizations have secondary, shorter timescales (down to
about three months), that may be related to the advection process of the magnetic field, or to the presence of several bright spots that compete amongst one another
for the scattering areas. 
Moreover, this behavior could also be modulated by radial pulsation modes, although no clear common timescales really emerges apart from the longest one.
Finally, rotation modulation, even if the latter is long, may also induce projection effects on the measured polarization.

Thus continuing long-term spectropolarimetric observations suggest a coherent picture in which bright spots trace the upflows of convection cells 
that are at the origin of the linear polarization, while the downflows of the same convection cells concentrate the observed magnetic field.
Such a dynamical picture should be closely linked to the multiple components \citep[e.g.,][]{jp07} of the surface velocity field.
Since pure convective motions seem insufficient to lead to the velocities observed for Antares \citep{owh17}, a magnetic component could help
to solve this problem.
Conversely, it is well known that a nonuniform distribution of the velocity field over the stellar surface may lead to strongly distorted profiles,
that coud affect somewhat the classical $S$ shape we used here.
Indeed, \citet{m88} showed that the first moments of both Stokes $I$ and $V$ profiles may be used for the purpose of setting constraints
on the magnetic field geometry.
As a first step, the characterization of the atmospheric dynamics will be described in a forthcoming paper.

\begin{acknowledgements}
We acknowledge the use of the VALD (Vienna, Austria) and Simbad (CDS, Strasbourg, France) databases. 
We thank the TBL team for providing service observing with NARVAL. 
We acknowledge financial support from ``Programme National de Physique Stellaire" (PNPS) of CNRS/INSU, France.
This project has received funding from the European Union's Horizon 2020 research and innovation program under the Marie Skodowska-Curie 
Grant agreement No. 665501 with the research Foundation Flanders (FWO) ([PEGASUS]$^2$ Marie Curie fellowship 12U2717N awarded to MM).
GAW acknowledges support in the form of a Discovery Grant from the Natural Sciences and Engineering Research Council (NSERC) of Canada.
Finally, the authors are grateful to the anonymous referee for useful remarks and comments.
\end{acknowledgements}

\setcounter{table}{0}
\onecolumn
\begin{longtable}{l c c c c l }
\caption{\label{tab1}Log of Stokes $V$ observations of \object{Betelgeuse} (for details, see Sect.\,2).} \\
\hline
\hline
Date               &  HJD         & Season & Instr. & n  & label \\
                   &(2 450 000 +) &        &        &    &       \\
\hline
\hline
\endfirsthead
\caption{continued.}\\
\hline\hline
Date               &  HJD         & Season & Instr. & n  & label \\
                   &(2 450 000 +) &        &        &    &       \\
\hline
\hline
\endhead
\hline
\hline
\endfoot
\hline
September 28, 2009 & 5103.120 & S1 &  E     &  2 &            \\
October 02, 2009   & 5107.024 & S1 &  E     &  1 &            \\
October 07, 2009   & 5112.038 & S1 &  E     &  3 &            \\
March 14, 2010     & 5270.397 & S1 &  N     & 16 & N20100300  \\
March 15, 2010     & 5271.373 & S1 &  N     & 16 & N20100300  \\
March 17, 2010     & 5273.312 & S1 &  N     & 16 & N20100300  \\
March 22, 2010     & 5278.362 & S1 &  N     & 16 & N20100300  \\
April 05, 2010     & 5292.334 & S1 &  N     & 20 & N20100400  \\
April 09, 2010     & 5296.319 & S1 &  N     & 20 & N20100400  \\
April 17, 2010     & 5304.341 & S1 &  N     & 19 & N20100400  \\
\hline
September 19, 2010 & 5459.690 & S2 &  N     & 16 &            \\ 
October 13, 2010   & 5483.706 & S2 &  N     & 16 &            \\
November 21, 2010  & 5521.895 & S2 &  E     & 12 & E20101121  \\
December 09, 2010  & 5540.668 & S2 &  N     & 16 & N20101209  \\
January 19, 2011   & 5581.433 & S2 &  N     & 16 &            \\
February 05, 2011  & 5598.381 & S2 &  N     & 16 & N20110205  \\
March 18, 2011     & 5639.369 & S2 &  N     & 16 &            \\
\hline
October 08, 2011   & 5843.710 & S3 &  N     & 16 & N20111008  \\
November 11, 2011  & 5877.659 & S3 &  N     & 16 & N20111111  \\
December 08, 2011  & 5904.436 & S3 &  N     & 16 & N20111208  \\
January 07, 2012   & 5934.573 & S3 &  N     & 16 &            \\
February 10, 2012  & 5968.373 & S3 &  N     & 16 &            \\
February 11, 2012  & 5969.454 & S3 &  N     & 16 &            \\
February 14, 2012  & 5971.952 & S3 &  E     & 11 &            \\
March 11, 2012     & 5998.377 & S3 &  N     & 16 &            \\
\hline
September 26, 2012 & 6197.026 & S4 &  E     & 11 & E20121000  \\
October 01, 2012   & 6202.035 & S4 &  E     & 11 & E20121000  \\
November 25, 2012  & 6257.172 & S4 &  E     & 16 & E20121200  \\
November 28, 2012  & 6260.164 & S4 &  E     & 16 & E20121200  \\
November 30, 2012  & 6262.041 & S4 &  E     &  7 &            \\
December 05, 2012  & 6266.984 & S4 &  E     & 23 & E20121200  \\
December 07, 2012  & 6269.113 & S4 &  E     & 11 & E20121200  \\
December 09, 2012  & 6271.136 & S4 &  E     & 11 & E20121200  \\
December 21, 2012  & 6282.874 & S4 &  E     & 11 &            \\
December 23, 2012  & 6285.087 & S4 &  E     & 22 &            \\
December 27, 2012  & 6288.937 & S4 &  E     & 11 & E20130000  \\
December 29, 2012  & 6290.949 & S4 &  E     & 11 & E20130000  \\
January 01, 2013   & 6293.988 & S4 &  E     & 11 & E20130000  \\
\hline
September 08, 2013 & 6544.693 & S5 &  N     & 16 & N20130908  \\
September 21, 2013 & 6557.660 & S5 &  N     & 16 & N20130921  \\
October   06, 2013 & 6572.700 & S5 &  N     & 10 & N20131006  \\
October   07, 2013 & 6573.705 & S5 &  N     &  6 & N20131006  \\
October   31, 2013 & 6597.709 & S5 &  N     & 16 & N20131031  \\
November  27, 2013 & 6624.600 & S5 &  N     & 12 & N20131127  \\
December  11, 2013 & 6638.613 & S5 &  N     & 12 &            \\
December  20, 2013 & 6647.510 & S5 &  N     & 12 &            \\
January   09, 2014 & 6667.517 & S5 &  N     & 12 &            \\
April     08, 2014 & 6756.334 & S5 &  N     & 14 & N20140408  \\
\hline
September 12, 2014 & 6913.668 & S6 &  N     & 16 & N20140900  \\
September 24, 2014 & 6925.647 & S6 &  N     & 16 & N20140900  \\
October   17, 2014 & 6948.697 & S6 &  N     & 16 & N20141000  \\
October   23, 2014 & 6954.597 & S6 &  N     & 16 & N20141000  \\
November  05, 2014 & 6967.611 & S6 &  N     & 16 &            \\
November  12, 2014 & 6974.564 & S6 &  N     & 16 &            \\
November  20, 2014 & 6982.634 & S6 &  N     & 16 &            \\
December  18, 2014 & 7010.578 & S6 &  N     & 16 & N20141218  \\
January   06, 2015 & 7029.446 & S6 &  N     & 16 & N20150106  \\
January   17, 2015 & 7040.388 & S6 &  N     & 12 &            \\
March     03, 2015 & 7085.304 & S6 &  N     & 16 &            \\
April     01, 2015 & 7114.310 & S6 &  N     & 16 & N20150401  \\
\hline
September 20, 2015 & 7286.676 & S7 &  N     & 16 &            \\
October   15, 2015 & 7311.712 & S7 &  N     & 16 &            \\
November  14, 2015 & 7341.706 & S7 &  N     & 16 &            \\
November  16, 2015 & 7343.582 & S7 &  N     & 16 &            \\
December  11, 2015 & 7368.592 & S7 &  N     & 16 &            \\
January   20, 2016 & 7408.537 & S7 &  N     & 16 & N20160120  \\
February  16, 2016 & 7435.344 & S7 &  N     & 16 &            \\
March     12, 2016 & 7460.354 & S7 &  N     & 16 &            \\
April     18, 2016 & 7497.341 & S7 &  N     & 16 & N20160418  \\
\hline
September 12, 2016 & 7644.682 & S8 &  N     & 16 & N20160912  \\
October   08, 2016 & 7670.694 & S8 &  N     & 16 & N20161008  \\
November  01, 2016 & 7694.561 & S8 &  N     & 16 & N20161101  \\
December  03, 2016 & 7726.580 & S8 &  N     & 16 & N20161203  \\
December  20, 2016 & 7743.479 & S8 &  N     & 16 &            \\
February  17, 2017 & 7802.397 & S8 &  N     & 16 & N20170217  \\
April     03, 2017 & 7847.319 & S8 &  N     &  8 & N20170403  \\
\hline                                  
\end{longtable}

\end{document}